\documentclass[twocolumn,aps,superscriptaddress,floatfix,prl]{revtex4}
\usepackage{mathrsfs}
\usepackage{amssymb}
\usepackage{amsmath}
\usepackage{graphicx}
\usepackage[normalem]{ulem}
\usepackage[dvips]{color}
\usepackage{bm}
\usepackage{longtable}
\usepackage{slashed}
\usepackage{epstopdf}

\usepackage{times}

\renewcommand\sout{\bgroup \color{red} \ULdepth=-.5ex \ULset}

\renewcommand{\rm}[1]{\textrm{#1}}

\newcommand{\ud}{{\rm{d}}}

\newcommand{\ue}{{\rm{e}}}

\newcommand{\keV}{{\ \rm{keV}}}
\newcommand{\fm}{{\ \rm{fm}}}

\begin{document}

\title{Neutron Skin in CsI and Low-Energy Effective Weak Mixing Angle from COHERENT Data}

\author{Xu-Run Huang}
\affiliation{School of
Physics and Astronomy and Shanghai Key Laboratory for Particle
Physics and Cosmology, Shanghai Jiao Tong University, Shanghai
200240, China}
\author{Lie-Wen Chen\footnote{%
Corresponding author: lwchen$@$sjtu.edu.cn}} \affiliation{School of
Physics and Astronomy and Shanghai Key Laboratory for Particle
Physics and Cosmology, Shanghai Jiao Tong University, Shanghai
200240, China}
\date{\today}

\begin{abstract}
Both the neutron skin thickness $\Delta R_{np}$ of atomic nuclei and the low-energy neutrino-nucleon ($\nu N$) interactions are of fundamental importance in nuclear and particle physics, astrophysics as well as new physics beyond the standard model (SM) but largely uncertain currently, and the coherent elastic neutrino-nucleus scattering (CE$\nu$NS) provides a clean way to extract their information. New physics beyond the SM may cause effectively a shift of the SM weak mixing angle $\theta_W$ in low-energy $\nu N$ interactions, leading to an effective weak mixing angle $\theta^*_W$. By analyzing the CE$\nu$NS data of the COHERENT experiment, we find that while a one-parameter fit to the COHERENT data by varying $\Delta R_{np}$ produces $\Delta R^{\rm{CsI}}_{np} \simeq 0.68^{+0.91}_{-1.13}$ fm for CsI with an unrealistically large central value by fixing $\sin^2 \theta^*_W$ at the low-energy SM value of $\sin^2\theta_W^{\rm{SM}} = 0.23857$, a two-dimensional fit by varying $\Delta R_{np}$ and $\sin^2 \theta^*_W$ leads to a strong positive correlation between $\Delta R_{np}$ and $\sin^2 \theta^*_W$ with significantly smaller central values of $\Delta R^{\rm {CsI}}_{np} \simeq 0.24_{-2.03}^{+2.30}$ fm and $\sin^2 \theta^*_W = 0.21_{-0.10}^{+0.13}$.
Although
the uncertainty is too large to claim a determination of $\Delta R^{\rm{CsI}}_{np}$ and $\sin^2 \theta^*_W$, the present study suggests that the multi-dimensional fit is important in future analyses of high-precision CE$\nu$NS data.
The implication of the possible deviation of $\sin^2 \theta^*_W$ from $\sin^2\theta_W^{\rm{SM}}$ on new physics beyond the SM is also discussed.
\end{abstract}

\maketitle

\textit{Introduction.}---
The neutron skin thickness of atomic nuclei, defined as
$\Delta R_{np}=R_n - R_p$ where $R_{n(p)} = \langle r_{n(p)}^{2}\rangle ^{1/2}$ is the neutron (proton)
rms radius of the nucleus, provides a good probe of the equation of state (EOS) for isospin asymmetric
nuclear matter~\cite{Bro00,Hor01,Fur02,Yos04,Che05b,Tod05,Cen09,Che10,Roc11,Agr12},
which is critically important due to its multifaceted roles in nuclear physics
and astrophysics~\cite{Lat04,Ste05,Bar05,LCK08} as well as some issues of new physics beyond
the standard model (SM)~\cite{Hor01b,Sil05,Wen09,Zhe14,Zhe15}.
While the $R_{p}$ can be measured precisely from electromagnetic processes (see, e.g., Refs.~\cite{Ang13,Fri95,Bla05}),
the $R_{n}$ is largely uncertain since it is usually determined from strong
processes, which is generally model dependent due to the complicated nonperturbative effects.
This provides a strong motivation for the Lead Radius Experiment (PREX)
being performed at the Jefferson Laboratory to determine the
$R_{n}$ of $^{208}$Pb to about $1\%$ accuracy by
measuring the parity-violating electroweak asymmetry in the elastic scattering
of polarized electrons from $^{208}$Pb~\cite{PREX}.
The PREX Collaboration reported the first result of the parity violating weak neutral
interaction measurement of the $\Delta R_{np}$ for $^{208}$Pb, i.e.,
$\Delta R^{208}_{np} = 0.33^{+0.16}_{-0.18}$ fm~\cite{Abr12}
(see, also, Ref.~\cite{Hor12}). The central value of $0.33$ fm
means a surprisingly large neutron skin thickness in $^{208}$Pb although
there is no compelling reason to rule out a such large value~\cite{Fat13}.

Recently, the COHERENT Collaboration~\cite{Aki17}
reported the first observation of the coherent elastic neutrino-nucleus
scattering (CE$\nu$NS)~\cite{Fre74,Fre77}.
In Ref.~\cite{Cad18a}, a value of the averaged $\Delta R_{np}$ of
$^{133}_{\ 55}$Cs and $^{127}_{\ 53}$I, i.e., $\Delta R^{\rm{CsI}}_{np} \simeq 0.7^{+0.9}_{-1.1} \fm$,
is extracted from analyzing the COHERENT
data. The extracted central value of $\Delta R^{\rm{CsI}}_{np} \simeq 0.7$ fm is
unrealistically large. To the best of our knowledge, $\Delta R^{\rm{CsI}}_{np} \simeq 0.7$ fm is actually
much larger than all the predictions of current nuclear models.
Moreover,
since $^{208}$Pb is much more neutron-rich than
$^{133}_{\ 55}$Cs and $^{127}_{\ 53}$I, the $\Delta R^{\rm{CsI}}_{np}$ is
expected to be smaller than the $\Delta R^{208}_{np}$ according to the neutron
skin systematics~\cite{Trz01,Swi05},
and thus $\Delta R^{\rm{CsI}}_{np} \simeq 0.7$~fm is inconsistent with
the PREX result.
Although the uncertainty is too large to claim a determination of $\Delta R^{\rm{CsI}}_{np}$,
the best-fit value $\Delta R^{\rm{CsI}}_{np} \simeq 0.7$~fm
indicates the possibility of a unrealistically large neutron skin thickness.
The possible inconsistency could be a hint of new physics in neutrino physics
and this provides the main motivation of the present work.

We note that in Ref.~\cite{Cad18a},
the $\Delta R^{\rm{CsI}}_{np}$ is extracted from a one-parameter fit to
the COHERENT data by varying $\Delta R^{\rm{CsI}}_{np}$ with the low-energy weak mixing
angle $\theta_W$ fixed at the SM value $\sin^2\theta_W^{\rm{SM}} = 0.23865$ obtained in
the modified minimal subtraction ($\overline{\rm{MS}}$) renormalization
scheme at near zero momentum transfer $Q=0$~\cite{Pat16} (the newest value is
$\sin^2\theta_W^{\rm{SM}} = 0.23857(5)$~\cite{Tan18}).
Experimentally,
the precise determination of $\sin^2\theta_W$ at low $Q^2$
is an ongoing issue~\cite{Kum13}, and the atomic parity violation (APV)
experiments offer the most precise results to date.
For example, by measuring the $6 s_{1/2} - 7 s_{1/2}$
electric dipole transition in $^{133}\rm{Cs}$ atom,
a value of $\sin^2 \theta_W = 0.2356(20)$
at $\left\langle Q \right\rangle \simeq 2.4$ MeV
is obtained~\cite{Por09,Dzu12,Rob15}, which is smaller than
$\sin^2\theta_W^{\rm{SM}}$ by about 1.5$\sigma$.
In the mid-energy regime,
the Qweak Collaboration reported the recent measurement
on proton's weak charge and obtained
$\sin^2 \theta_W = 0.2383(11)$ at $Q = 0.158$ GeV~\cite{And18},
agreeing well with the SM prediction.
On the other hand, the low-energy neutrino-nucleon ($\nu N$) interactions
could involve new physics beyond the
SM~\cite{Kum13,Bil18,Bar05b,Lin17,Lia17,Pap18,Giu15,Cad18b,Dav12},
which may cause effectively a shift of
the SM weak mixing angle $\theta_W$ in the $\nu N$ interactions, leading
to a low-energy effective weak mixing angle $\theta^*_W$.
Any experimental constraints on $\theta^*_W$ would provide useful
information on new physics beyond the SM.

In this work, we extract the values of $\Delta R^{\rm{CsI}}_{np}$ and
$\sin^2 \theta^*_W$ using a two-dimensional (2D) fit to the COHERENT data
by varying $\Delta R^{\rm{CsI}}_{np}$ and $\sin^2 \theta^*_W$.
Compared to the results
using one-parameter fit with $\sin^2 \theta^*_W$ fixed at $\sin^2\theta_W^{\rm{SM}}$,
we find a strong positive correlation between $\Delta R_{np}$ and $\sin^2 \theta^*_W$
with significantly smaller central values of
$\Delta R^{\rm {CsI}}_{np} \simeq 0.24_{-2.03}^{+2.30}$ fm and $\sin^2 \theta^*_W = 0.21_{-0.10}^{+0.13}$
at $Q \simeq 0.05$ GeV (corresponding to the energy scale of COHERENT experiment),
indicating that the $\sin^2 \theta^*_W$ may play an important role in extracting
neutron skin information from analyzing the CE$\nu$NS data.

\textit{CE$\nu$NS in the COHERENT experiment.}---
The differential cross section for coherent elastic neutrino-nucleus scattering has a
straightforward SM prediction in the case with different proton and neutron distributions
(form factors) in the nucleus. By neglecting the radiative corrections and axial contributions,
the cross section can be expressed as~\cite{Bar05b,Lin17,Pat12,Can18,Sch18}:
\begin{align}
\label{XSection}
\dfrac{\ud \sigma}{\ud T} (E_\nu, T) &= \dfrac{G^2_F M}{2 \pi} G_V^2 \Bigg[ 1 - \dfrac{M T}{E^2_\nu}  + \Big( 1 - \dfrac{T}{E_\nu} \Big)^2 \Bigg] , \\
\label{Gv}
G_V &=  Z g^p_V F_p(q^2) + N g^n_V F_n(q^2)  ,
\end{align}
where $G_F$ is the Fermi coupling constant, $M$ is the nucleus mass,
$E_\nu$ and $T$ are neutrino energy and nuclear recoil kinetic energy,
respectively.
For a given $E_\nu$, the corresponding $T$ varies from $0$ to
$T^{max} = 2 E^2_\nu / (M + 2 E_\nu)$.
The proton and neutron neutral current vector couplings are defined,
respectively, as $g_V^p = \frac{1}{2} - 2 \sin^2 \theta_W$ and
$g_V^n = - \frac{1}{2}$.
The form factor $F_{n(p)}(q^2)$ encapsulate the neutron (proton)
number density distribution in nuclei, where the momentum transfer
$q$ is given by $q^2 = 2 E_\nu^2 T M / (E_\nu^2 - E_\nu T) \simeq 2 M T$
under the condition of $E_\nu \gg T$.

In the case of the COHERENT experiment, the measurement is performed using
a CsI detector which is dominantly composed of
$^{133}_{\ 55}$Cs and $^{127}_{\ 53}$I.
The mass of a nucleus with $N(Z)$ neutrons (protons) is determined by its
corresponding total binding energy ($E_B$) from $M = N\times m_n + Z \times m_p - E_B$
where $m_{n(p)}$ is the rest mass of
neutrons (protons).
The binding energies per nucleon are $8.40998$ MeV and $8.44549$ MeV~\cite{Wan17} for
isotopes $^{133}\rm{Cs}$ and $^{127}\rm{I}$, respectively.
As for their density distributions, in order to test the model dependence,
two analytic nuclear form factors are adopted, namely, the symmetrized Fermi (SF)
form factor and the Helm form factor, which are two very successful and well-tested
forms of nuclear form factors for medium to heavy nuclei~\cite{Hel56,De87,Has88,Cad18a,Pie16}.
Both form factors are characterized by two parameters related to
the nuclear radius and the surface thickness (diffuseness), respectively.

The SF form factor has the form (See, e.g., Ref.~\cite{Pie16})
\begin{eqnarray}\label{SFermiFF}
F_{\rm{SF}}(q^2) &=& \dfrac{3}{q c[(q c)^2 + (\pi q a)^2]} \Bigg[ \dfrac{\pi q a}{\sinh (\pi q a)} \Bigg] \notag\\
&& \times \Bigg[ \dfrac{\pi q a}{\tanh (\pi q a)} \sin (q c) - q c \cos (q c) \Bigg],
\end{eqnarray}
and the corresponding rms radius is expressed as
\begin{equation}\label{RrmsSF}
R_{\rm{SF}}^2 \equiv \left\langle r^2 \right\rangle = \dfrac{3}{5} c^2 + \dfrac{7}{5} (\pi a)^2 .
\end{equation}
where $c$ is the half-density radius and $a$ quantifies the surface thickness
$t = 4 a \ln 3$. Experimentally, the proton distribution has been determined
precisely, and we take the same parameters for
proton distribution as in Ref.~\cite{Cad18a}, which are obtained by fitting the
proton structure data of $^{133}\rm{Cs}$ and $^{127}\rm{I}$ measured in muonic atom
spectroscopy, namely, $t_p = 2.30 \fm$, $c_{p,\rm{Cs}} = 5.6710 \pm 0.0001 \fm$ and
$c_{p,\rm{I}} = 5.5931 \pm 0.0001 \fm$. The corresponding proton rms radii for
$^{133}\rm{Cs}$ and $^{127}\rm{I}$ are $R^{\rm{Cs}}_p = 4.804$ fm
and $R^{\rm{I}}_p = 4.749$ fm, respectively.

The Helm form factor is expressed as~\cite{Hel56}
\begin{equation}\label{HelmFF}
F_{\rm{Helm}}(q^2) = 3 \dfrac{j_1(q R_0)}{q R_0} \ue^{- q^2 s^2 /2} ,
\end{equation}
where $j_1(x)$ is the spherical Bessel function of order one, i.e.,
$j_1(x) = \sin(x)/x^2 - \cos(x)/x$. The rms radius is simply given by
\begin{equation}\label{RrmsHelm}
R_{\rm{Helm}}^2 \equiv \left\langle r^2 \right\rangle = \dfrac{3}{5} R_0^2 + 3 s^2,
\end{equation}
where $R_0$ is the box radius and $s$ quantifies the surface thickness.
Again, for the proton distributions in $^{133}\rm{Cs}$ and $^{127}\rm{I}$,
we use $s_p = 0.9$ fm following Ref.~\cite{Cad18a},
which was determined for the proton form factor of similar nuclei~\cite{Fri82},
and the $R_{0,p}$ is determined by the corresponding $R_p$.

For the parameters of the neutron distributions in $^{133}\rm{Cs}$ and
$^{127}\rm{I}$, they are essentially unknown. In these neutron-rich nuclei,
in principle, the neutron distributions should be different from the proton
distributions because of the charge difference, which means that the neutron
distributions could have different radius parameters ($c_n$ and $R_{0,n}$)
and diffuseness (surface thickness) parameters ($t_n$ and $s_n$) compared to the proton
distributions. We will examine these effects in the following.

In the COHERENT experiment, the potoelectrons are counted to monitor the
scattering events and extract the nuclear recoil energy, with approximately
$1.17$ photoelectrons expected per keV of nuclear recoil energy, denoted as
$\zeta = 1.17 \keV^{-1}$~\cite{Aki17}.
The number of event counts in a nuclear recoil energy bin
$[T^i, T^{i+1}]$ can be obtained as
\begin{eqnarray}
N_i^{th} &=& N_{\rm{CsI}} \sum_{\nu_l} \sum_{\cal{N} = \rm{Cs},\rm{I}} \int_{T^i}^{T^{i+1}} \ud T \mathcal{A}(\zeta T) \notag\\
&& \times \int_{E_{\nu}^{min}}^{E_{\nu}^{max}} \ud E_{\nu} \dfrac{\ud N_{\nu_{l}}}{\ud E_\nu} \dfrac{\ud \sigma_{\nu-\cal{N}}}{\ud T} ,
\end{eqnarray}
where $N_{\rm{CsI}}$ is the number of CsI in the detector and
is given by $N_A m_{\rm {det}}/M_{\rm{CsI}}$ with
$N_A$ being the Avogadro constant, $m_{\rm {det}} = 14.57$ kg  the detector mass
and $M_{\rm{CsI}} = 259.8\ \rm{g}/\rm{mol}$ the molar mass of CsI.
The acceptance efficiency function $\mathcal{A}(x)$ is decribed by~\cite{Aki18}
\begin{equation}
\begin{split}
\mathcal{A}(x) = \dfrac{a}{1 + \ue^{- k (x - x_0)}} \Theta(x-5),
\end{split}
\end{equation}
where the parameter values are taken as $a = 0.6655^{+0.0212}_{-0.0384}$,
$k = 0.4942^{+0.0335}_{-0.0131}$ and $x_0 = 10.8507^{+0.1838}_{-0.3995}$,
and the $\Theta(x)$ is a modified Heaviside step function
defined as
\begin{equation}
\begin{split}
\Theta (x - 5) =
\begin{cases}
0 & x < 5, \\
0.5 & 5 \leq x < 6, \\
1 & x \geq 6 .
\end{cases}
\end{split}
\end{equation}
The value of $E^{min}_\nu$ depends on $T$, and the $E^{max}_\nu$ is
related to the neutrino source.
At the Spallation Neutron Source, the neutrino flux is generated from
the stopped pion decays $\pi^+ \rightarrow \mu^+ + \nu_\mu$ as well as
the subsequent muon decays $\mu^+ \rightarrow e^+ + \bar{\nu}_\mu + \nu_e$.
The neutrino population has the following energy
distributions~\cite{Cad18a,Lia17}
\begin{equation}
\begin{split}
\dfrac{\ud N_{\nu_\mu}}{\ud E_\nu} &= \eta \delta\Bigg( E_\nu - \dfrac{m_\pi^2 - m_\mu^2}{2 m_\pi} \Bigg) , \\
\dfrac{\ud N_{\bar{\nu}_\mu}}{\ud E_\nu} &= \eta \dfrac{64 E_\nu^2}{m_\mu^3} \Bigg( \dfrac{3}{4} - \dfrac{E_\nu}{m_\mu} \Bigg) , \\
\dfrac{\ud N_{\nu_e}}{\ud E_\nu} &= \eta \dfrac{192 E_\nu^2}{m_\mu^3} \Bigg( \dfrac{1}{2} - \dfrac{E_\nu}{m_\mu} \Bigg) ,
\end{split}
\end{equation}
with $E^{max}_\nu \leq m_\mu /2$.
The normalization factor $\eta$ is defined as
$\eta = r N_{\rm{POT}}/(4 \pi L^2)$, where $r = 0.08$ is the averaged
production rate of the decay-at-rest (DAR) neutrinos for each flavor
per proton on target, $N_{\rm{POT}} = 1.76 \times 10^{23}$ is the total
number of protons delivered to the target and
$L = 19.3 \ \rm{m}$ is the distance between the neutrino source and
the CsI detector~\cite{Aki17}.

To evaluate the fitting quality on the COHERENT data in Fig. 3A of
Ref.~\cite{Aki17}, following Ref.~\cite{Cad18a},
we apply the following least-squares function with only
the $12$ energy bins from $i = 4$ to $i = 15$, i.e.,
\begin{eqnarray}
\chi^2 &=& \sum_{i=4}^{15} \Bigg( \dfrac{N_i^{exp} - (1 + \alpha)N_i^{th} - (1 + \beta)B_i}{\sigma_i} \Bigg)^2 \notag \\
&& +\Bigg( \dfrac{\alpha}{\sigma_\alpha} \Bigg)^2 + \Bigg( \dfrac{\beta}{\sigma_\beta} \Bigg)^2 .
\end{eqnarray}
Here for each energy bin, the experimental number of events,
denoted as $N^{exp}_i$, is generated from the C-AC differences,
and $B_i$ is the estimated beam-on background with only prompt
neutrons included~\cite{Aki17}.
The $\sigma_i = \sqrt{N^{exp}_i + 2 B_i^{ss} + B_i}$ is the statistical
uncertainty where $B^{ss}_i$ is the estimated steady-state background
determined with AC data~\cite{Aki17}.
The $\alpha$ and $\beta$ are the systematic parameters corresponding to the
uncertainties on the signal rate and the beam-on background rate, respectively.
The fractional uncertainties corresponding to 1-$\sigma$ variation are
$\sigma_\alpha = 0.28$ and $\sigma_\beta = 0.25$~\cite{Aki17}.
All the experimental data are taken from the COHERENT release~\cite{Aki18}.

\textit{Results and discussions.}---
In the present work for CE$\nu$NS calculations,
we replace the $\theta_W$ in Eq.~(\ref{XSection}) by $\theta^*_W$
to effectively consider the possible effects of new physics
in $\nu N$ interactions.
We first assume that the neutron and proton distributions have the same
diffuseness parameters (i.e., $t_n = t_p$ and $s_n = s_p$) and
the value of $\sin^2 \theta^*_W$ is fixed at the SM value of
$\sin^2 \theta_W^{\rm{SM}} = 0.23857$, and then perform a one-parameter
fit to the COHERENT data by varying $R_n$ to extract the neutron rms
radius $R^{\rm{CsI}}_n$ of CsI
($^{133}_{\ 55}$Cs and $^{127}_{\ 53}$I are assumed to have equal $R_n$).
Our calculations lead to $R_n = 5.46^{+0.91}_{-1.13}$ fm with the
Helm form factor and $R_n = 5.47^{+0.91}_{-1.13}$ fm with the SF form factor.
Our results thus nicely confirm the value
of $R_n = 5.5^{+0.9}_{-1.1}$ fm extracted in Ref.~\cite{Cad18a}
with the same assumptions.

In addition, we explore the effects of the neutron diffuseness
parameters. To this end, we perform a one-parameter fit to the COHERENT data
by varying $R^{\rm{CsI}}_n$ with various fixed values of the diffuseness parameter
while the effective weak mixing angle is fixed at $\sin^2 \theta^*_W = \sin^2 \theta_W^{\rm{SM}}$.
The results indicate that a variation of $\pm 0.02 \fm$ for $\Delta R^{\rm{CsI}}_{np}$
arises when $s_n$ changes from $0.63$ to $1.17 \fm$
(corresponding to a variation of $\pm 30\%$ for $s_n=0.9$ fm) in the Helm form factor.
The same conclusion is obtained when the SF form factor is used.
Therefore, compared to the obtained neutron skin thickness of
$\Delta R^{\rm{CsI}}_{np} \simeq 0.68^{+0.91}_{-1.13}$ fm,
the effects of the neutron diffuseness parameters are indeed
quite small, consistent with the statement in
Ref.~\cite{Cad18a}.

\begin{figure}[htbp]
\centering
\includegraphics[width = 0.45\textwidth]{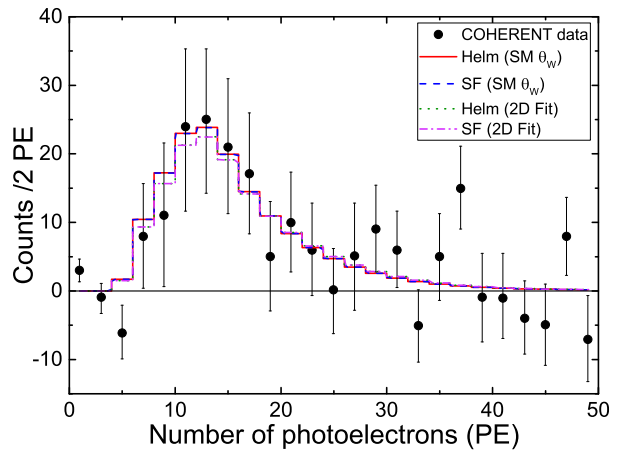}
\caption{(Color online) The CE$\nu$NS event counts as a function of the number of
photoelectrons in the COHERENT experiment.
The solid (dashed) line shows the results with best-fit
neutron rms radius using the Helm (SF) form factor in the one-parameter
fit when the $\sin^2 \theta^*_W$ is fixed at the SM prediction.
The dotted (dash-dotted) line shows the results with best-fit parameters
in the 2D fit using the Helm (SF) form factor.
Data are taken from Ref.~\cite{Aki17}.}
\label{CEvNSCounts}
\end{figure}

\begin{figure}[htbp]
\centering
\includegraphics[width = 0.4\textwidth]{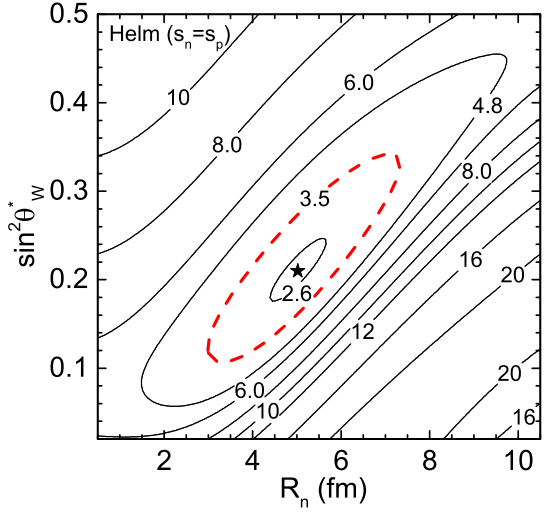}
\caption{(Color online) The $\chi^2$ contours in the plane of $R_n$ vs $\sin^2 \theta^*_W$
obtained from a 2D fit to the COHERENT data using the Helm form factor with $s_n = s_p$.
The star marks the center values of $R_n = 5.02 \fm$ and $\sin^2 \theta^*_W = 0.21$
at $\chi^2_{\rm{min}} = 2.498$. The dashed curve corresponds to the contour at
$\chi^2 = \chi^2_{\rm{min}} + 1$.}
\label{Chi2Helm}
\end{figure}

Now we turn to examining the effects of the low-energy effective weak mixing angle.
The possible non-standard running of $\sin^2 \theta^*_W$ in
low-energy regime is expected to influence the extraction of the neutron
distribution from the low-energy CE$\nu$NS experiments.
The simultaneous precise determination of the neutron distribution and
the low-energy $\sin^2 \theta^*_W$ through CE$\nu$NS experiments can (in)validate
our knowledge of nuclear physics and neutrino physics.
Hence, we perform a 2D fit to the COHERENT data by varying $R_n$ and
$\sin^2\theta^*_W$ using the Helm form factor with $s_n = s_p$.
The resulting number of CE$\nu$NS event counts as a function of the number of
photoelectrons is shown in Fig.~\ref{CEvNSCounts}
while the corresponding $\chi^2$ contours
are displayed in Fig.~\ref{Chi2Helm}.

For comparison, we also include in Fig.~\ref{CEvNSCounts} the corresponding
results from the COHERENT data, the similar 2D fit by using the SF
form factor with $t_n = t_p$, and the one-parameter fit by varying $R_n$
with fixed $\sin^2 \theta^*_W = \sin^2 \theta_W^{\rm{SM}}$ using both the Helm
and SF form factors.
It is seen from Fig.~\ref{CEvNSCounts} that for both one-parameter
and 2D fits, the SF and Helm form factors produce almost identical
results, indicating the independence
of our results on the form of nuclear form factors.
Furthermore, Fig.~\ref{CEvNSCounts} indicates that compared to the one-parameter
fit, the 2D fit predicts a fewer event counts in the region of $7 \sim 15$
for the photoelectron number, leading to a decreases by $\sim 3.2\%$ for
the number of total event counts.

From Fig.~\ref{Chi2Helm}, one sees clearly that there exhibits a strong positive
correlation between $R_n$ and $\sin^2\theta^*_W$.
Particularly interesting is that there exists favored center
values for $R_n$ and $\sin^2\theta^*_W$, i.e.,
\begin{equation}
	\begin{split}
	R_n^{\rm{Helm}} = 5.02_{-2.03}^{+2.30} \fm, \ \sin^2 \theta^*_W = 0.21_{-0.10}^{+0.13}.
	\end{split}
\end{equation}
We note that very similar results are obtained when the SF form factor
is used.
With the averaged rms radii of protons and neutrons in
$^{133}\rm{Cs}$ and $^{127}\rm{I}$, we then obtain the
averaged neutron skin thickness of CsI as
\begin{equation}
	\Delta R^{\rm {CsI}}_{np} \simeq 0.24_{-2.03}^{+2.30} \fm.
\end{equation}
The favored central value $\Delta R^{\rm {CsI}}_{np} \simeq 0.24$~fm is significantly
smaller than $\Delta R^{\rm {CsI}}_{np} \simeq 0.68$~fm
extracted from the one-parameter fit to the COHERENT data with fixed
$\sin^2 \theta^*_W = \sin^2 \theta_W^{\rm{SM}}$,
indicating the importance of the $\sin^2 \theta^*_W$ in the extraction of
$\Delta R^{\rm {CsI}}_{np}$ from CE$\nu$NS.

Furthermore, we examine the effects of neutron diffuseness
parameters using the 2D fit to the COHERENT data by varying $R_n$ and
$\sin^2 \theta^*_W$ with $s_n$ and $t_n$ fixed at various values.
Our results indicate that the central value of $\Delta R^{\rm {CsI}}_{np}$ varies by
$\pm 0.03$ fm (the corresponding $R_n$ varies from $4.99$ fm to $5.05$ fm)
when the value of $s_n$ in the Helm form factor changes from
$0.63$~fm to $1.17$~fm (corresponding to a variation of
$\pm 30\%$ for $s_n=0.9$ fm).
Similarly, we find the central value of $\Delta R^{\rm {CsI}}_{np}$ varies by
$\pm 0.04$ fm (the corresponding $R_n$ varies from $4.99$ fm to $5.07$ fm)
when the value of $t_n$ in the SF form factor
changes from $1.61$~fm to $2.99$~fm (corresponding to a variation of
$\pm 30\%$ for $t_n=2.3$ fm).
Meanwhile, we note the central value variation of $\sin^2 \theta^*_W$ is tiny, namely,
from $0.209$ to $0.211$ when $s_n$ ($t_n$) changes from $0.63$ ($1.61$)~fm to $1.17$ ($2.99$)~fm.
The variation of $\pm (0.03\sim 0.04)$ fm is appreciable compared to the central
value $\Delta R^{\rm {CsI}}_{np} \simeq 0.24$ fm, implying that one may extract
useful information on the neutron diffuseness parameters
in atomic nuclei from analyzing the future high-precise data of CE$\nu$NS via
a three-dimensional fit by varying $\sin^2 \theta^*_W$, $\Delta R^{\rm {CsI}}_{np}$
and the diffuseness parameters ($s_n$ and $t_n$).
This can help to address the interesting question about whether the neutron skin structure is really from
the bulk radius difference or the surface diffuseness difference between the neutron and
proton distributions in atomic nuclei~\cite{Trz01,War10,Tar14}.
Therefore, our results suggest that a multi-dimensional fit
is important to extract the value of $\sin^2 \theta^*_W$ and the neutron skin information
including its size (i.e., $\Delta R^{\rm {CsI}}_{np}$) and shape (e.g., $s_n$ and $t_n$)
in future analyses of high-precision CE$\nu$NS data.
Nevertheless, the extracted central value of
$\Delta R^{\rm {CsI}}_{np} \simeq 0.24$ fm with an uncertainty of $\pm (0.03\sim 0.04)$
fm obtained in the present work is consistent with some carefully calibrated
nuclear models (see, e.g., Refs.~\cite{Cad18a,Fat13}).

On the other hand, a possible substantial deviation of $\sin^2 \theta^*_W$ from
$\sin^2\theta_W^{\rm{SM}}$, i.e., $\Delta \sin^2 \theta^*_W = - 0.02857$,
is obtained with the best-fit value of $\sin^2 \theta^*_W = 0.21$.
This deviation could be a hint of new physics beyond SM in neutrino physics.
For example,
one new physics scenario is to introduce the nonstandard interactions (NSIs) in the SM interactions,
which has been widely discussed~\cite{Bar05b,Lin17,Lia17,Bil18,Pap18}.
To make a rough estimate on the parameters in NSIs, we introduce an ad hoc
nonstandard charge $G_V^{\rm{NSI}}$ to replace the $G_V$ in
Eq.~(\ref{Gv}), i.e.,
\begin{eqnarray}\label{DF-GVNSI}
G_V^{\rm{NSI}} &=& Z g_V^p F_p (q^2) + N g_V^n F_n (q^2) \notag \\
&&+ 3 \delta_{\rm{NSI}} [ Z F_p (q^2) + N F_n (q^2) ],
\end{eqnarray}
where $\delta_{\rm{NSI}} = \epsilon^{uV}_{\alpha \alpha} = \epsilon^{dV}_{\alpha \alpha}$ ($\alpha = e,\mu,\tau$ represents the neutrino flavor)
denotes the NSI parameters.
Eq.~({\ref{DF-GVNSI}}) can be obtained from the more general NSIs
(see, e.g., Refs.~\cite{Bil18,Bar05b,Lin17,Pap18}) by neglecting the flavor-changing
couplings $\epsilon^{qV}_{\alpha \beta}$ ($\alpha \neq \beta$) and assuming that
the new flavor-preserving couplings ($\epsilon^{qV}_{\alpha \alpha}$)
are flavor symmetric for neutrinos and the first-generation quarks ($q=u,d$).
Then one can estimate the value of $\delta_{\rm{NSI}}$ as
\begin{equation}\label{NSI}
	\begin{split}
	\delta_{\rm{NSI}} \simeq - \frac{2 Z}{3 A} \Delta \sin^2 \theta^*_W = 0.008 ,
	\end{split}
\end{equation}
by assuming $F_p (q^2) \simeq F_n (q^2)$.
These results indicate that the NSI contribution into the proton and neutron neutral
current vector couplings is $3 \delta_{\rm{NSI}} = 0.024$, which is even
larger than the SM proton coupling $g_V^p = \frac{1}{2} - 2 \sin^2 \theta^{\rm{SM}}_W = 0.02286$.

Moreover, we would like to point out that the deviation of
$\sin^2 \theta^*_W$ from $\sin^2\theta_W^{\rm{SM}}$
in neutrino physics can also potentially arise from the neutrino
electromagnetic properties, e.g.,
the neutrino charge radius $\left\langle r^2_\nu \right\rangle$~\cite{Giu15,Pap18,Cad18b}.
Furthermore, the deviation could be as well from the dark parity
violation~\cite{Kum13,Dav12}.
All these scenarios beyond the SM can effectively shift the low-energy
weak mixing angle in $\nu N$ interactions and worthy of further investigation with
forthcoming more precise CE$\nu$NS data in future.
It will be also very interesting to check the similar effects in
other weak neutral interaction measurements, e.g., APV and PREX.

Finally, it should be pointed out that the uncertainty of the extracted
values for both $\Delta R^{\rm{CsI}}_{np}$ and $\sin^2 \theta^*_W$ is very large
due to the poor statistics of the current COHERENT data, and this hinders us from
claiming a determination of the $\Delta R^{\rm{CsI}}_{np}$ and $\sin^2 \theta^*_W$.
Nevertheless, our results indicate that the $\Delta R^{\rm{CsI}}_{np}$ is positively correlated
with $\sin^2 \theta^*_W$ and the best-fit values lead to the possibility of significantly smaller
values of $\Delta R^{\rm{CsI}}_{np}$ and $\sin^2 \theta^*_W$ compared to the one-parameter fit
to the COHERENT data with $\sin^2\theta^*_W = \sin^2\theta_W^{\rm{SM}}$.
The present work thus suggests that
the $\sin^2 \theta^*_W$
may play an important role in extracting neutron skin information from analyzing
the CE$\nu$NS data and
a multi-dimensional fit is important
in future analyses of high-precision CE$\nu$NS data.

\textit{Summary and outlook.}---
We have demonstrated that the low-energy effective weak mixing angle $\theta^*_W$
plays an important role in the extraction of neutron skin information of atomic nuclei
from the CE$\nu$NS experiments.
By analyzing the CE$\nu$NS data of the COHERENT experiment,
we have found that
while a one-parameter fit to the COHERENT data produces
$\Delta R^{\rm{CsI}}_{np} \simeq 0.68^{+0.91}_{-1.13}$ fm
with $\sin^2\theta^*_W = \sin^2\theta_W^{\rm{SM}} = 0.23857$,
a two-dimensional fit by varying $\Delta R_{np}$ and $\sin^2 \theta^*_W$
leads to a strong positive correlation
between $\Delta R_{np}$ and $\sin^2 \theta^*_W$
with significantly smaller central values of
$\Delta R^{\rm {CsI}}_{np} \simeq 0.24_{-2.03}^{+2.30}$ fm
and $\sin^2 \theta^*_W = 0.21_{-0.10}^{+0.13}$.
While the best-fit value $\Delta R^{\rm{CsI}}_{np} \simeq 0.24$ fm
seems to be reasonable, the substantial deviation of the best-fit value
$\sin^2 \theta^*_W  = 0.21$ from $\sin^2\theta_W^{\rm{SM}}$ could give a
hint on new physics in $\nu$-nucleon interactions.

Although the current large uncertainty does not allow
us to claim a determination of the $\Delta R^{\rm{CsI}}_{np}$ and $\sin^2 \theta^*_W$,
our present work suggests that a multi-dimensional fit is important to
extract useful information on neutron skin information (including its size and shape)
and the low-energy effective
$\sin^2 \theta^*_W$ from analyzing the high-precision data of future CE$\nu$NS measurements.
It will be also extremely interesting
to explore the similar effects in other experiments of
weak neutral interaction measurements.

\textit{Acknowledgments.}---
We thank Juan I. Collar for the useful communication on the quenching factor for CsI,
Xiao-Gang He and Alexander I. Studenikin for useful discussions.
This work was supported in part by the National Natural Science
Foundation of China under Grant No. 11625521, the Major State Basic Research
Development Program (973 Program) in China under Contract No.
2015CB856904, the Program for Professor of Special Appointment (Eastern
Scholar) at Shanghai Institutions of Higher Learning, Key Laboratory
for Particle Physics, Astrophysics and Cosmology, Ministry of
Education, China, and the Science and Technology Commission of
Shanghai Municipality (11DZ2260700).





\end{document}